\PassOptionsToPackage{hyphens}{url}
\documentclass[%
reprint,
superscriptaddress,
amsmath,amssymb,
aps,longbibliography,
floatfix,
]{revtex4-1}

\usepackage{graphicx}
\usepackage{verbatim}
\usepackage[version=4]{mhchem}
\usepackage{xcolor}
\usepackage{physics}
\usepackage[normalem]{ulem}
\usepackage{pdfpages}
\usepackage{pgffor}
\usepackage{orcidlink}
\usepackage{xcolor}

\usepackage{url}
\usepackage{xcolor}

\setcitestyle{numbers,square}

\makeatletter
\AtBeginDocument{\let\LS@rot\@undefined}
\makeatother

\begin{document}

\title{A strategic roadmap for an atomistic machine-learning ecosystem}

\author{J\"org Behler \orcidlink{0000-0002-1220-1542}}
\affiliation{Lehrstuhl f\"ur Theoretische Chemie II, Ruhr-Universit\"at Bochum, 44780 Bochum, Germany}
\affiliation{Research Center Chemical Sciences and Sustainability, Research Alliance Ruhr, 44780 Bochum, Germany}
\author{Michele Ceriotti \orcidlink{0000-0003-2571-2832}}
\email{michele.ceriotti@epfl.ch}
\affiliation{Laboratory of Computational Science and Modeling, Institut des Matériaux, École Polytechnique Fédérale de Lausanne, 1015 Lausanne, Switzerland}
\author{Cecilia Clementi \orcidlink{0000-0001-9221-2358}}
\affiliation{Department of Physics, Freie Universit\"at Berlin, 14195 Berlin, Germany}
\author{G\'abor Cs\'anyi \orcidlink{0000-0002-8180-2034}}
\affiliation{Engineering Laboratory, University of Cambridge, Trumpington St, Cambridge, UK}
\affiliation{Max Planck Institute for Polymer Research, Ackermannweg 10, Mainz, Germany}
\author{Alin-Marin Elena \orcidlink{0000-0002-7013-6670}}
\affiliation{Scientific Computing Department, Science and Technology Facilities Council, Daresbury Laboratory, Keckwick Lane, Daresbury WA4 4AD, UK}
\author{Aditi Krishnapriyan \orcidlink{0000-0003-3472-6080}}
\affiliation{Lawrence Berkeley National Laboratory (LBNL)}
\affiliation{University of California, Berkeley}

\author{Joseph W. Abbott \orcidlink{0000-0002-0502-6790}}
\affiliation{Laboratory of Computational Science and Modeling, Institut des Matériaux, École Polytechnique Fédérale de Lausanne, 1015 Lausanne, Switzerland}
\author{Fabio Affinito \orcidlink{0000-0003-0716-2849}}
\affiliation{CINECA, Consorzio Interuniversitario, Bologna, Italy}
\author{Albert P. Bart\'ok \orcidlink{0000-0002-4347-8819}}
\affiliation{Department of Physics, University of Warwick, Coventry, CV4 7AL, UK}
\affiliation{Warwick Centre for Predictive Modelling, School of Engineering, University of Warwick, Coventry, CV4 7AL, UK}
\author{Ilyes Batatia \orcidlink{0000-0001-6915-9851}}
\affiliation{Engineering Laboratory, University of Cambridge, Trumpington St, Cambridge, UK}
\author{Filippo Bigi \orcidlink{0000-0002-9338-7317}}
\affiliation{Laboratory of Computational Science and Modeling, Institut des Matériaux, École Polytechnique Fédérale de Lausanne, 1015 Lausanne, Switzerland}
\affiliation{FAIR, Meta, San Francisco, CA, USA}
\author{Florian N. Br\"unig \orcidlink{0000-0001-8583-6488}}
\affiliation{Department of Physics and Materials Science, University of Luxembourg, L-1511 Luxembourg}
\author{Yannick Calvino Alonso~\orcidlink{0009-0008-9573-7772}}
\affiliation{Laboratory for Computational Molecular Design, Institute of Chemical Sciences and Engineering,
\'{E}cole Polytechnique F\'{e}d\'{e}rale de Lausanne, 1015 Lausanne, Switzerland}
\author{Giuseppe Carleo \orcidlink{0000-0002-8887-4356}}
\affiliation{Laboratory of Computational Quantum Science, Institut de Physique, École Polytechnique Fédérale de Lausanne, 1015 Lausanne, Switzerland}
\author{Aur\'elie Champagne \orcidlink{0000-0002-6013-2887}}
\affiliation{Université de Bordeaux, CNRS, Bordeaux INP, ICMCB, F-33600 Pessac, France}
\affiliation{Réseau sur le Stockage Electrochimique de l’Energie (RS2E), CNRS FR 3459, Cedex 1, F-80039 Amiens, France}
\author{Stefan Chmiela \orcidlink{0000-0003-0892-952X}}
\affiliation{Machine Learning Group, Technische Universit\"at Berlin, 10587 Berlin, Germany}
\affiliation{Berlin Institute for the Foundations of Learning and Data -- BIFOLD, 10587 Berlin, Germany}
\author{Marc L. Descoteaux \orcidlink{0000-0001-8504-3429}}
\affiliation{John A. Paulson School of Engineering and Applied Sciences,
Harvard University, Cambridge, MA, USA}
\author{Ralf Drautz \orcidlink{0000-0001-7101-8804}} 
\affiliation{Interdisciplinary Centre for Advanced Materials Simulation (ICAMS), Ruhr-University Bochum, 44780  Bochum, Germany}
\author{Alexandra Farcas \orcidlink{0000-0001-9972-9376}}
\affiliation{Department of Physics and Chemistry, Technical University of Cluj-Napoca, 400641 Cluj-Napoca, Romania}
\affiliation{
National Institute for Research and Development of Isotopic and Molecular Technologies, 67-103 Donat, 400293 Cluj-Napoca, Romania}
\author{Meng Gao \orcidlink{0009-0004-1439-3308}}
\affiliation{FAIR, Meta, San Francisco, CA, USA}
\author{Rohit Goswami \orcidlink{0000-0002-2393-8056}}
\affiliation{Laboratory of Computational Science and Modeling, Institut des Matériaux, École Polytechnique Fédérale de Lausanne, 1015 Lausanne, Switzerland}
\author{Michael F. Herbst \orcidlink{0000-0003-0378-7921}}
\affiliation{Mathematics for Materials Modelling, Institute of Mathematics \& Institute of Materials,  École Polytechnique Fédérale de Lausanne, 1015 Lausanne, Switzerland}
\author{Christian Holm
\orcidlink{0000-0003-2739-310X}}
\affiliation{Institute for Computational Physics, University of Stuttgart, Stuttgart, Germany}
\author{James R. Kermode\orcidlink{0000-0001-6755-6271}}
\affiliation{Warwick Centre for Predictive Modelling, School of Engineering, University of Warwick, Coventry, CV4 7AL, UK}
\author{Alexander L. M. Knoll \orcidlink{0009-0007-1754-031X}}
\affiliation{Lehrstuhl f\"ur Theoretische Chemie II, Ruhr-Universit\"at Bochum, 44780 Bochum, Germany}
\affiliation{Research Center Chemical Sciences and Sustainability, Research Alliance Ruhr, 44780 Bochum, Germany}
\author{Tobias Kreiman \orcidlink{0009-0002-1142-5744}}
\affiliation{University of California, Berkeley}
\author{Marcel F. Langer \orcidlink{0000-0002-1270-3016}}
\affiliation{Laboratory of Computational Science and Modeling, Institut des Matériaux, École Polytechnique Fédérale de Lausanne, 1015 Lausanne, Switzerland}
\author{Hoang-Thien Luu \orcidlink{0009-0008-0961-1891}}
\affiliation{Institute of Metallurgy, Technical University of Clausthal, Clausthal-Zellerfeld, Germany}
\author{Yury Lysogorskiy \orcidlink{0000-0003-4617-3188}} 
\affiliation{Interdisciplinary Centre for Advanced Materials Simulation (ICAMS), Ruhr-University Bochum, 44780  Bochum, Germany}
\author{Mihai-Cosmin Marinica\orcidlink{0000-0002-3994-6771}}
\affiliation{Universit\'{e} Paris-Saclay, CEA, Service de recherche en Corrosion et Comportement des Mat\'{e}riaux, SRMP, 91191, Gif-sur-Yvette, France}
\author{Rocco Meli \orcidlink{0000-0002-2845-3410}}
\affiliation{Swiss National Supercomputing Centre, ETH Zürich, Lugano, Switzerland}
\author{Klaus-Robert M\"{u}ller\orcidlink{0000-0002-3861-7685}}
\affiliation{Machine Learning Group, Technische Universit\"at Berlin, 10587 Berlin, Germany}
\affiliation{Berlin Institute for the Foundations of Learning and Data -- BIFOLD, 10587 Berlin, Germany}
\affiliation{Max Planck Institute for Informatics, Stuhlsatzenhausweg, 66123 Saarbr{\"u}cken, Germany}
\affiliation{Department of Artificial Intelligence, Korea University, Anam-dong, Seongbuk-gu, 02841 Seoul, Korea}
\author{Frank Noé \orcidlink{0000-0003-4169-9324}}
\affiliation{Microsoft Research AI for Science, 10178 Berlin, Germany}
\affiliation{Department of Mathematics, Freie Universit\"at Berlin, 14195 Berlin, Germany}
\author{Mohamadhosein Nosratjoo\orcidlink{0000-0003-2036-1586}}
\affiliation{Department of Chemistry, The University of Manchester, Manchester, UK}
\author{Simon Olsson\orcidlink{0000-0002-3927-7897}}\affiliation{Department of Computer Science and Engineering, Chalmers University of Technology and University of Gothenburg, SE-41296 Gothenburg, Sweden}
\author{Christoph Ortner\orcidlink{0000-0003-1498-8120}}
\affiliation{Department of Mathematics, University of British Columbia, Vancouver, Canada}
\author{Aldo S. Pasos-Trejo
\orcidlink{0000-0002-4462-2835}}
\affiliation{Department of Physics, Freie Universit\"at Berlin, 14195 Berlin, Germany}
\author{Anyang Peng \orcidlink{0000-0002-0630-2187}}
\affiliation{AI for Science Institute, Beijing, China}
\author{Eric Qu \orcidlink{0000-0001-8816-869X}}
\affiliation{University of California, Berkeley}
\author{Andrea Rizzi \orcidlink{0000-0001-7693-2013}}
\affiliation{Achira Inc., San Francisco, California, USA}
\author{Mariana Rossi \orcidlink{0000-0002-3552-0677}}
\affiliation{MPI for the Structure and Dynamics of Matter, Hamburg, Germany}
\affiliation{Yusuf Hamied Department of Chemistry, Cambridge University, Cambridge, UK}
\author{Bassem Sboui \orcidlink{0009-0006-5465-2110}}
\affiliation{Université de Bordeaux, CNRS, Bordeaux INP, ICMCB, F-33600 Pessac, France}
\affiliation{Réseau sur le Stockage Electrochimique de l’Energie (RS2E), CNRS FR 3459, Cedex 1, F-80039 Amiens, France}
\author{Gregor N.\ C.\ Simm\,\orcidlink{0000-0001-6815-352X}}
\affiliation{Microsoft Research, AI for Science, Cambridge CB1\,2FB, UK}
\author{Alexandre Tkatchenko \orcidlink{0000-0002-1012-4854}}
\affiliation{Department of Physics and Materials Science, University of Luxembourg, L-1511 Luxembourg}

\author{Jacopo Venturin \orcidlink{0009-0003-7347-3036}}
\affiliation{Department of Physics, Freie Universit\"at Berlin, 14195 Berlin, Germany}
\author{O. Anatole von Lilienfeld \orcidlink{0000-0001-7419-0466}}
\affiliation{Department of Chemistry, Department of Materials Science \& Engineering, and Department of Physics, University of Toronto, Toronto, Ontario, Canada}
\affiliation{Vector Institute for Artificial Intelligence, Toronto, Ontario, Canada}
\author{William C. Witt \orcidlink{0000-0002-1578-1888}}
\affiliation{John A. Paulson School of Engineering and Applied Sciences,
Harvard University, Cambridge, MA, USA}
\author{Brandon M. Wood \orcidlink{0000-0001-6295-7535
}}
\affiliation{FAIR, Meta, San Francisco, CA, USA}
\author{Tigany Zarrouk \orcidlink{0000-0001-9396-0163}}
\affiliation{Department of Chemistry and Materials Science, Aalto University, 02150 Espoo, Finland}
\author{Fabian Zills \orcidlink{0000-0002-6936-4692}}
\affiliation{Institute for Computational Physics, University of Stuttgart, Stuttgart, Germany}

\date{\today}

\begin{abstract}
Data-driven machine learning (ML) techniques have become an essential tool in many domains of science. Their application to atomistic simulations of matter is particularly widespread and impactful. This success is due largely to the existence of a well-developed and established physics-based modeling framework, ranging from first-principles electronic-structure calculations to molecular dynamics and statistical sampling, into which ML was integrated naturally to reshape long-standing trade-offs between accuracy, efficiency, and scale.
Nevertheless, this integration raises both conceptual and practical challenges, from choosing between data-centric and physics-based modeling approaches to adapting established software stacks to modern hardware accelerators and ML libraries.
As the field evolves rapidly, fueled in part by widespread enthusiasm but also by tangible impact, it seems appropriate to take a moment to consider the current state of the art and open challenges, and reflect on what can be done to better coordinate efforts across the community.   
With this goal in mind, several members of this community met in Lausanne in January 2026 at CECAM to discuss algorithms, models, software and hardware infrastructure, and the most promising scientific applications that have become possible thanks to the use of artificial intelligence in atomic-scale simulations. This strategic roadmap paper summarizes the outcomes of these discussions, suggesting some long-term goals, and some concrete actions, to establish a healthy, sustainable and impactful atomistic ML ecosystem.

\end{abstract}

\maketitle

\section{What is atomistic machine learning} 
\label{sec:intro}

The simulation of matter at the atomic scale is one of the most established and impactful applications of computing in science and engineering. Decades of work on electronic-structure theory, statistical sampling, and molecular dynamics (MD) have produced a mature framework that, in many cases, can not only explain but also effectively complement or even anticipate experimental findings~\cite{Peng2017,Chen2012,Bowman2015,P3817}.
Yet, the cost of accurate first-principles calculations limits what can actually be simulated to much smaller systems and shorter timescales than what is needed to address the most consequential questions in chemistry, biology, and materials science. In the last two decades, machine learning (ML) has emerged as the most successful strategy to lift these limitations, and the speed at which it is doing so is striking enough to deserve a coordinated reflection on where the field is heading.

This roadmap builds on the discussions of a workshop held at CECAM in Lausanne in January 2026\footnote{CECAM (Centre Européen de Calcul Atomique et Moléculaire) is a European organization that hosts workshops in the area of atomic-scale simulations of matter. The web page of the workshop can be consulted at \protect\url{https://www.cecam.org/workshop-details/a-roadmap-for-an-atomistic-machine-learning-software-ecosystem-1376}.}, where developers and users of atomistic machine-learning methods came together to discuss algorithms, models, software and hardware infrastructure, as well as applications. Our scope is intentionally focused. We consider only \emph{atomistic} machine learning in the sense of particle-based modeling of matter, with the goal of predicting the structure and properties of matter by computer simulation from fundamental physical laws (meaning without experimental input), on the basis of quantum-chemical calculations, mainly density functional theory (DFT) but occasionally also higher-accuracy methods. We do not cover sequence-based generative models for biology in the style of AlphaFold~\cite{abra+24nature}, cheminformatics, or analysis pipelines that postprocess simulation outputs.

Atomistic ML is an application of ML in which three features are especially prominent. First, the underlying ``source of truth'' or physical framework is well understood, meaning that large amounts of training data can be generated almost routinely using electronic-structure calculations. Second, sharing data is harder than it should be: small differences in pseudopotentials, exchange–correlation functionals, basis sets, or convergence settings, which can all be perfectly justified on a per-project basis, often lead to numerically incompatible datasets. Third, the usual tension between interpolation and extrapolation is not a side issue but a central methodological question.

Reference datasets for a specific system of interest are used to construct an efficient machine-learning interatomic potential (MLIP), also termed a machine-learning force field (MLFF)---the two expressions originate from the materials-science and biochemical-modeling communities~\cite{Lifson1968}, respectively, and are used interchangeably for ML models; we use the former in this roadmap. MLIPs are then used to scale up simulations both in size and time, to gain insights inaccessible to electronic-structure calculations, which requires a reliable description clearly beyond the geometries covered in the training set. Additionally, models are often expected to describe chemistries that are not present in their training sets, i.e., potentials should be applicable to systems with notably different structures or chemical composition. Unless there is reliable extrapolation beyond the training domain, the model is not useful.

The community now finds itself with a powerful new set of tools in a fast-evolving but uneven landscape. Methods, datasets, and software stacks are proliferating; new models appear almost weekly, and benchmarks struggle to keep up. We argue that this is precisely the moment to coordinate. The aim of this roadmap is to summarize the state of the art and the most pressing open problems, and to identify some community actions that we believe would have a strong impact on the long-term health of the field. Sec.~\ref{sec:usecases} surveys the main use cases of atomistic ML, with an emphasis on interatomic potentials while also taking a deliberate look at other atomic properties and electronic-structure surrogates. Sec.~\ref{sec:beyond-props} discusses approaches to reach longer time and length scales: coarse-grained models, enhanced sampling, generative samplers, and learned integrators. Sec.~\ref{sec:sota} discusses current models, datasets, and representative applications. Sec.~\ref{sec:assess} addresses the assessment of model and data quality, including benchmarks, challenges, and uncertainty quantification. Sec.~\ref{sec:swhw} examines software and hardware considerations. Sec.~\ref{sec:strategy} closes with some concrete recommendations.

\section{Use cases for atomistic machine learning}
\label{sec:usecases}

 \begin{figure}
    \centering\includegraphics[width=1.0\linewidth]{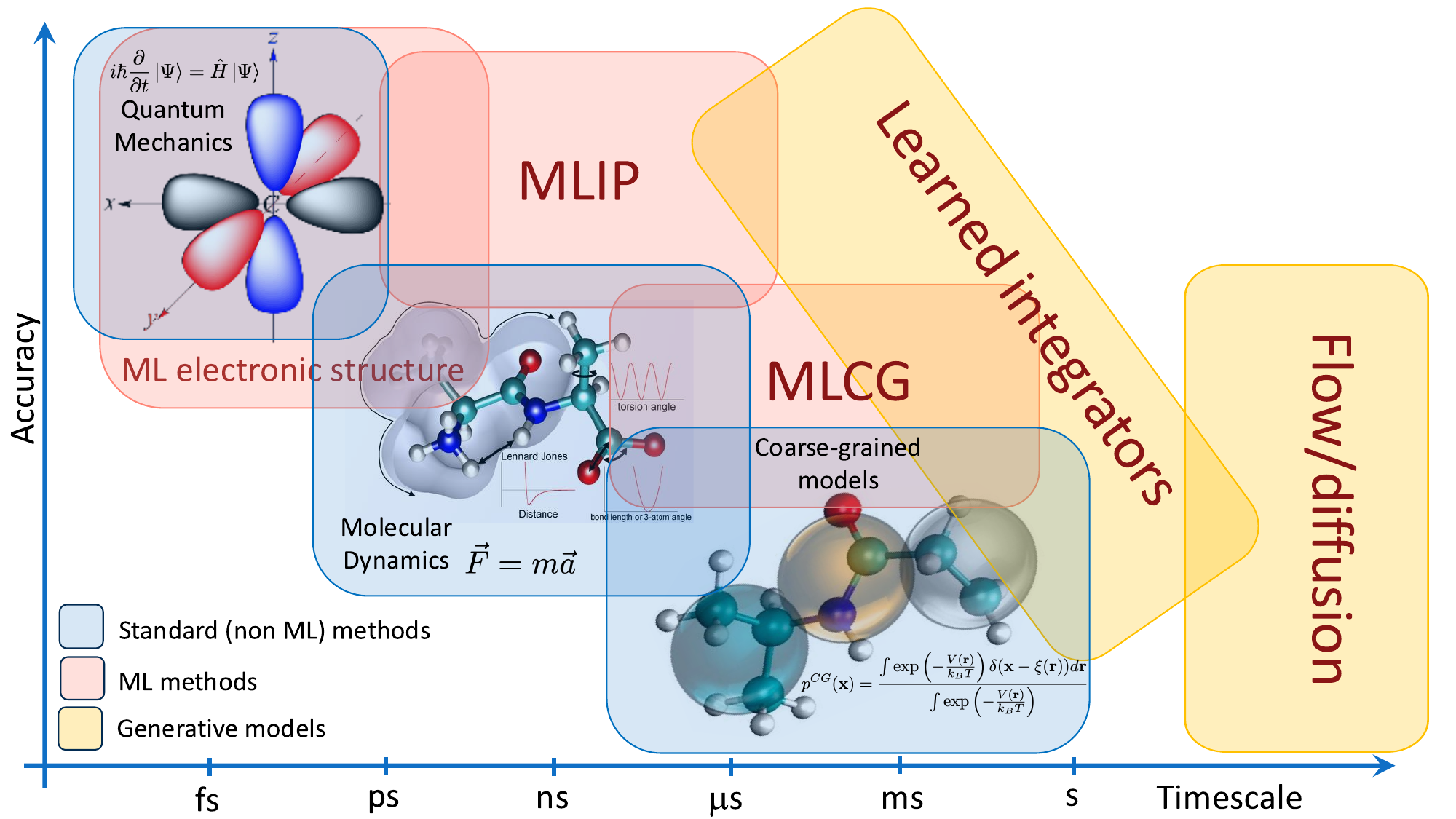}
    \caption{Accuracy and timescales accessible for traditional approaches and atomistic machine learning.}
    \label{fig:approaches}
\end{figure}

By far the most mature application of atomistic machine learning is the parameterization of interatomic potentials, and it is what most of this roadmap concentrates on. The reach of atomistic ML, however, is much broader, and several adjacent problems are evolving fast enough to be discussed alongside interatomic potentials rather than in isolation. We sketch the landscape here, leaving a more critical discussion of the state of the art to Sec.~\ref{sec:sota}.

\subsection{Machine learning interatomic potentials and force fields}
\label{sec:mlip}

Interatomic potentials are multidimensional functions that map a configuration of atoms (and, where applicable, the vectors defining a periodic unit cell) to its potential energy and the associated forces and stresses. Their construction has been a central problem of computational physics and chemistry since their inception, motivated by the wish to reach the accuracy of electronic-structure calculations for molecular dynamics and geometry optimization applications but without the cost of representing the electrons explicitly. The first ML approaches that were transferable across configurations of a given system were introduced for adsorbed molecules in the mid-1990s~\cite{blan+95jcp}. Broad adoption at that time was still hindered by the restriction of early MLIPs to only a few degrees of freedom. The modern field can be traced back to the introduction of atom-centered geometry representations exploiting locality~\cite{behl-parr07prl}, which enabled the construction of potentials for very large systems. The earliest examples for material systems were high-dimensional neural-network potentials (HDNNPs)~\cite{behl-parr07prl} based on atom-centered symmetry functions~\cite{behl11jcp}, and shortly afterwards Gaussian Approximation Potentials (GAP) based on spherical harmonic representations and kernel regression~\cite{bart+10prl,bart+13prb}---well before the deep-learning boom in other fields of science. 
Kernel and neural models were also shown to be capable of learning quantum energies and related properties across chemical compound space, with Coulomb-matrix representations being applied to QM7/QM9-type molecular data~\cite{rupp+12prl,mont+12nips,rama+14sd}, initiating a rapid development that quickly reduced errors relative to DFT below chemical accuracy and, with $\Delta$-learning, brought predictions close to chemical accuracy relative to higher levels of theory~\cite{rama+15jctc}.
MLIPs have evolved significantly since then, going through successive generations~\cite{deri+19am,behler2021four,deri+21cr,mlff-chemrev}, improving the range of applications, the complexity of the underlying model, and the understanding of relationships between structural representations and architectures~\cite{pozd+20prl,musi+21cr,niga+22jcp2,P6787}. 
From those foundations, the field has expanded into a dense ecosystem of diverse architectures and software stacks that we discuss in Sec.~\ref{sec:sota}.

Two regimes have emerged over the past few years. On one side, models trained for a specific system or chemistry can reach excellent accuracy, rivaling electronic structure calculations with modest training sets (on the order of a few thousand structures). This regime is relevant whenever a study requires very high quantitative fidelity, high efficiency, or both, and is well served by both shallow and deep architectures. On the other side, increasingly large and chemically diverse datasets have enabled the training of \emph{general-purpose} potentials, sometimes referred to as ``universal'' or ``foundation'' models, that aim to cover a substantial fraction of the periodic table with a single set of learned parameters~\cite{chen-ong22ncs,deng2023chgnet,yang2024mattersimdeeplearningatomistic,barrosoluque2026openmaterials2024omat24,neumann2024,rhodes2025orbv3atomisticsimulationscale,zhang2025graphneuralnetworkera,batatia2025crosslearningelectronicstructure,BatatiaPOLAR1,malosso2026high,li2026dpa4pushingaccuracycostfrontier,Kim2026,lyso2026,merc+23nature,bata+25jcp,mazi+25ncomm, wood2026umafamilyuniversalmodels,park2024scalableparallelalgorithmgraph, yin2025alphanetscalinglocalframebasedatomistic, koker2025trainingfoundationmodelmaterials, xu2026spectralspatialtensoratomiccluster, zhou2026matrisreliableefficientpretrained, ho2026equivariantmanybodymessagepassing, liao2026equiformerv3scalingefficientexpressive, xu2026edgeclusterexpansionradial, kavanagh2026fastaccuratefoundationmodels, yan2025materialsfoundationmodelhybrid, yzchen08_eqnorm, fu25h, Zhang2024}. 
The terminology is not without controversy. ``Foundation'' is borrowed from natural language processing, where it denotes models pre-trained on broad data and then adapted to many downstream tasks. In this sense the label fits: general-purpose potentials are frequently fine-tuned, often with little data, to reach the accuracy required for a given system. The same need for fine-tuning, however, shows that ``universal'' is aspirational: current models are better described as broadly transferable energy and force predictors, with emerging extensions to derived properties. It is also worth keeping in mind that for molecular systems even early MLIPs boasted higher transferability than empirical potentials~\cite{smit+17cs,schu+17ncomm}, and that it is not difficult to find quantitative and qualitative failure modes for most general-purpose models~\cite{kreiman2026understanding}.
The change of the past few years is one of scale and accessibility: a researcher starting on a new system can now obtain a reasonable first structural model with little or no in-house training---without the parameterization effort of a classical force field, and at a fraction of the cost of a first-principles simulation---which has simplified the workflow of computational studies and has also made the fast screening of large numbers of molecules or materials practical, boosting data-driven materials discovery~\cite{RoadmapDataMaterialsBauer_2024}.

The two regimes are not in opposition. General-purpose models can serve as initial guesses to be specialized through fine-tuning~\cite{bata+25jcp, Radova.2025,lyso2026,koker2026pftphononfinetuningmachine}, distillation~\cite{amin2025towards}, or retraining on a focused dataset. Specialized models continue to set the standard of accuracy for individual systems and, increasingly, also serve as cheap inference targets for production simulations. In many cases, however, fine-tuning is superfluous, given that the zero-shot error of universal models against their reference electronic-structure model when performing sophisticated simulation tasks is often lower than the error of the reference against experiments~\cite{mazi+25ncomm}. This also means that further improvements depend as much on the quality of the training data as on advances in ML itself.
This roadmap proposes that the community think about training data, software interfaces, and benchmarks in a way that reflects the continuum from specialized to generally applicable models.

\subsection{What are MLIPs good for}
\label{sec:mlip-uses}

MLIPs can be used in all applications for which an empirical potential or an \emph{ab initio} ground-state potential calculation could be used, but with higher accuracy than the former, and greater speed than the latter~\cite{deri+19am,behler2021four,Jia_2020}. 
The most impactful use case is probably that of MD---used to compute both equilibrium thermodynamic averages~\cite{chen+19pnas} and time-dependent behavior~\cite{zhou+23ne}---but similar acceleration can be exploited for high-throughput campaigns to look for locally stable geometries~\cite{merc+23nature} as well as to look for transition states~\cite{Wander_2025}.
For this latter task, which probes the regions of the potential energy surface for which it is hardest to collect training structures (and for which electronic structure calculations are least reliable), it might be advisable not to rely entirely on MLIPs, but to use them to accelerate transition-state search algorithms such as the dimer method, or the nudged elastic band family of methods~\cite{henkelmanDimerMethodFinding1999,jonssonNudgedElasticBand1998}.
Pairing MLIP-based optimization and \emph{ab initio} refinement can cut by up to 90\%{} the cost of the search of both local minima and saddle points by reducing the number of \emph{ab initio} force evaluations~\cite{Garijo_del_R_o_2019,goswamiEfficientImplementationGaussian2025}, while still delivering nominal first-principles accuracy.

\subsection{Other atomic and electronic properties}
\label{sec:other-props}
Many practical questions require properties beyond energies and forces. Multipole moments, polarizabilities, nuclear magnetic resonance (NMR) chemical shifts, x-ray photoelectron spectra (XPS) core-electron binding energies, infrared and Raman intensities, and other simple response properties have been targets of ML approaches for at least a decade~\cite{P2391,rupp+15jpcl,P6484,bart+17sa,veit+20jcp,keith2021combining,vonlilienfeld2020exploring,drau20prb,golze2022,Briling2024,montavon2013machine}, and are now mature enough to be routinely included as auxiliary outputs of MLIP architectures with little overhead. Such properties allow for not only the validation of generated structural models by comparison with experimental data but also the direct inference of atomistic structures that agree with the experiment using inverse methods~\cite{zarrouk2024,zarrouk2025,luong2026a}.

A more fundamental development is the prediction of electronic-structure outputs, such as the electron density, the density of states, the single-particle Hamiltonian, or the density matrix in a chosen basis~\cite{snyd+12prl,broc+17nc,bogojeski2020quantum, schu+19nc,chan+19npjcm,benm+20prb,west-maur21cs, kong2022, niga+22jcp, zhan+22npjcm, li+22ncs, gong+23nc, zhon+23npjcm, Qian2026, kaniselvan2026learning,Fabrizio2020}. 
These approaches are advancing quickly and represent a qualitatively different proposition: rather than replacing the electronic-structure potential energy surface with a surrogate, one replaces (parts of) the electronic-structure calculation itself, in a way that preserves access to all derived properties. 
This approach promises to reduce the computational cost of certain bottlenecks of the \textit{ab initio} calculation or even reduce the scaling with system size.
Most of these models regress Hamiltonian or density-matrix elements in a fixed atomic-orbital basis, inheriting its conventions rather than learning basis-independent observables directly, a choice that binds the target to an arbitrary representation and limits the opportunity for ML to drive model order reduction.
Intermediate approaches that combine ML with semi-empirical or tight-binding Hamiltonians retain an explicit electronic structure at low cost~\cite{zubatiuk2021machine,wang2021tightbinding,panosetti2020learning,fedik2023synergy}.
For these electronic-structure surrogates, the data, software, and validation challenges are correspondingly larger than for MLIPs; we return to them in Sec.~\ref{sec:swhw}. We expect this area to grow into a major component of the ecosystem in the coming years, and to enable a step change in the achievable accuracy of derived simulations.

\section{Longer time and length scales}
\label{sec:beyond-props}

\subsection{Coarse-graining}
\label{sec:cg}
Atomistic MLIPs hold the promise of near-quantum accuracy at a fraction of first-principles cost, greatly extending the time and length scales accessible to simulations (cf. Fig.~\ref{fig:approaches}), yet they remain more expensive than the classical force fields used in biomolecular simulations, where reaching beyond microsecond timescales is often necessary. 
Coarse-grained (CG) models introduce a second level of dimensional reduction: instead of replacing electronic structure by an atomistic potential, they emulate an atomistic simulation by an even lower-resolution model. A coarse-graining map projects atomistic coordinates onto CG coordinates, e.g., the positions of backbone and side-chain CG particles (or ``beads'') for an amino acid chain, or a small collection of beads representing the functional groups of molecules in the condensed phase. The formally exact target is the free energy surface, or equivalently the potential of mean force, whose Boltzmann distribution equals the marginal distribution obtained by mapping the atomistic equilibrium ensemble to the CG variables~\cite{Izvekov2005MSCG,Noid2008MSCG,Shell2008RelativeEntropy,Jin2022BottomUpCG}.
CG models accelerate simulations by reducing the number of particles that need to be treated for a given system size, and by allowing for longer time steps, since fast molecular motions such as the vibration of covalent bonds are averaged out.

Unlike atomistic MLIPs, machine-learned coarse-grained (MLCG) potentials are not generally trained on directly available pointwise energy labels. Systematic bottom-up approaches instead optimize CG force fields using variational force matching or relative-entropy minimization~\cite{Izvekov2005MSCG,Noid2008MSCG,Shell2008RelativeEntropy}. The exact CG potential is generally many-body, even when the underlying atomistic force field contains mostly few-body terms. This is where pre-ML CG force fields were limited in accuracy and transferability~\cite{clementi2008}, and where ML makes an important difference: flexible neural, kernel, graph neural network and cluster-expansion representations can approximate many-body CG free-energy surfaces and thereby produce bottom-up force fields that better reproduce atomistic equilibrium distributions~\cite{John2017GAPCG,Wang2019CGnets,Wang2020,Husic2020CGGNN,Wang2025ACECG,durumeric2023,kramer2023,durumeric2026}.

Bottom-up matching is not the only route to a CG model and is arguably not the most established one. Indeed, a large body of widely used CG force fields is parameterized top-down, by tuning interactions to reproduce experimental or macroscopic observables---partition free energies, densities, interfacial tensions---rather than the statistics of a finer simulation~\cite{peter2009multiscale,souza2021martini3}. The two philosophies have complementary failure modes: bottom-up models are systematically tied to, and limited by, the accuracy of their atomistic or even quantum-mechanical reference data, whereas top-down models can match target observables while misrepresenting the underlying ensemble. For ML, this suggests a direction that neither tradition has fully taken advantage of: hybrid objectives in which a force-matched or relative-entropy base model is corrected, regularized, or fine-tuned against experimental targets~\cite{liebl2025hybrid, navarro2023topdown}. Such hybrid training is also one of the few principled ways to escape the ``quantum ceiling'' discussed in Sec.~\ref{sec:uncertainty}, since the correction is anchored to experiment rather than to a reference method that may itself be wrong.

Recent MLCG work has moved from system-specific models toward transferable biomolecular force fields. Early neural and Gaussian-process CG models showed that force-matched ML potentials can reproduce fine-grained equilibrium distributions and folding free-energy landscapes for small molecular systems~\cite{John2017GAPCG,Wang2019CGnets,Husic2020CGGNN}. Protein-scale MLCG models now preserve important thermodynamic features of atomistic simulations while accelerating sampling by several orders of magnitude~\cite{Majewski2023ProteinThermodynamics}. The newest transferable protein MLCG models show that, with sufficiently diverse atomistic training data and expressive architectures, a single bottom-up force field can generalize across sequences and reproduce folded structures, intermediates, intrinsically disordered ensembles, folding-upon-binding, and mutation-induced stability changes~\cite{charron2025}. 

Together, these results point toward foundation-style CG simulators for proteins, polymers, membranes, and molecular assemblies---but realizing this ambition depends not only on expressive architectures and diverse atomistic training data~\cite{charron2025} but also on community agreement on standard mappings, shared CG datasets, and CG-appropriate benchmarks, the same coordination actions recommended for atomistic data in Sec.~\ref{sec:strategy}.

Several features set the evaluation of CG models apart from the atomistic case discussed in Sec.~\ref{sec:assess}. There are no pointwise energy labels to fit against; the natural reference quantities are distributional (CG-resolution radial distribution functions, free-energy surfaces, folding landscapes) or dynamical, and the matched forces are noisy gradients of a potential of mean force rather than deterministic targets. Uncertainty quantification for CG free-energy surfaces is correspondingly less developed than for atomistic MLIPs. CG also inherits, in sharper form, the data-compatibility problem that we stress for electronic-structure settings (Sec.~\ref{sec:data}): data generated under different mappings or at different state points cannot be naively assembled, because the underlying CG potential is itself mapping- and state-point-dependent~\cite{dunn2016vdw}. 

Important limitations remain. Transferability across thermodynamic conditions and to non-protein systems is an active research objective~\cite{Venturin2026}, and the CG mapping itself fixes what physics is retained.

The mapping---the number and location of the retained degrees of freedom---need not be fixed by hand, and could itself be
learned. Most current CG work assumes a
chemically motivated mapping and learns only the corresponding free energy surface. Yet the
mapping determines what physics is representable, and the retained
coordinates do not need to coincide with atomistic or molecular positions:
they could be a center of mass, a center of charge, or any other
collective coordinate. Learning the optimal map for a desired number of
reduced degrees of freedom, under an explicit accuracy or
information-preservation criterion, is therefore a well-posed problem in
its own right. Data-driven approaches to select a mapping---autoencoder
\cite{wang2019cgae} and graph-based~\cite{webb2019gbcg} constructions, or
variational criteria that optimize the mapping jointly with the model
\cite{giulini2020infotheory, foley2015resolution}---are a natural way
to address it, and the interaction between mapping choice, achievable
 accuracy for equilibrium properties, and dynamical correctness remains largely
unexplored~\cite{yang2023slicing}.
Thermodynamic consistency, even when achieved, does not deliver correct dynamics. Integrating out fast degrees of freedom removes friction and memory that the retained coordinates experienced implicitly, so a CG model that reproduces the atomistic equilibrium distribution will, in general, reproduce neither its diffusion constants nor its transition rates~\cite{Nuske2019}. Recovering physically meaningful timescales requires treating lost degrees of freedom explicitly, for example, through generalized Langevin or Mori–Zwanzig formulations with learned memory kernels~\cite{klippenstein2021memory}, or by learning the propagator rather than the potential. The latter connects coarse-graining directly to the learned stochastic integrators and implicit-transfer-operator models of Sec.~\ref{sec:force-free}~\cite{schr+23nips}: where a CG potential plus a thermostat cannot be expected to produce the correct kinetics, a learned transition density at CG resolution may be obtained, but at the cost of giving up the explicit energy function. Whether thermodynamically consistent CG potentials and kinetically faithful CG propagators can be unified in a single model is one of the central open problems.

CG simulations need to be further processed: their configurations are typically interpreted, analyzed, or re-coupled to finer descriptions, all of which require reconstructing atomistic detail from CG coordinates. This backmapping step is itself a conditional generative problem and can be addressed with the same flow- and diffusion-based machinery used for equilibrium sampling~\cite{2303.01569,2201.12176,shmilovich2022backmapping}, making decoding a natural companion to the encoding implied by the CG mapping. Treating mapping, potential, sampler, and backmapping as parts of one learnable pipeline---rather than as separate, manually bridged stages---is a promising direction that current MLCG efforts only partially realize.

\subsection{Enhanced sampling with ML potentials}
ML potentials accelerate energy and force evaluations compared with quantum chemistry methods, but still suffer from the sampling problem. Many target observables are ensemble or path quantities---free-energy differences, phase equilibria, nucleation events, reaction rates, conformational change and binding or folding populations---and direct trajectories can remain trapped in metastable basins and lead to intractable computational cost for direct MD simulation even when each force evaluation only takes milliseconds of wall-clock time~\cite{Olsson2026}. A key opportunity is therefore to combine MLIPs with established enhanced-sampling estimators such as umbrella sampling~\cite{Torrie1977Umbrella}, replica exchange~\cite{Sugita1999ReplicaExchange}, metadynamics~\cite{Laio2002Metadynamics}, thermodynamic integration~\cite{Kirkwood1935TI}, free-energy perturbation~\cite{Zwanzig1954FEP}, nonequilibrium switching~\cite{Jarzynski1997Nonequilibrium,Crooks1999Fluctuation}, and related methods~\cite{Henin2022EnhancedSampling}.

Metadynamics, and its evolution On-the-fly Probability Enhanced Sampling (OPES)~\cite{10.1021/acs.jpclett.0c00497}, have been recently combined with active learning to build reactive neural-network potentials and to obtain free-energy profiles for solution and catalytic chemistry, including urea decomposition in water and several Haber--Bosch-relevant surface reactions~\cite{Yang2022UreaMetadynamics,Bonati2023DynamicsHeterogeneousCatalysis,Tripathi2024PoisoningHaberBosch,Perego2024DataEfficientCatalysis,ToselloGardini2025BaH2Ammonia}, and enhanced sampling with MLIPs has been applied at scale to electrified catalytic interfaces~\cite{sahoo2026insightsdimerizationelectrifiedcu}. In condensed-phase chemistry, umbrella sampling has been combined with a neural network potential for studying Strecker synthesis~\cite{P7215}, alchemically equipped MLIPs have been used with rigorous free-energy perturbation protocols to compute solvation free energies with high accuracy~\cite{Moore2026SolvationMACE}, and implicit-solvent ML potentials have been combined with FEP/BAR-style path reweighting to accelerate solvation-free-energy estimation~\cite{Roecken2024ReSolv}. In materials science, MLIPs have been used in an on-the-fly Bayesian adaptive biasing force method for computing formation free energies in metastable basins~\cite{zhong+prm2023, zhong+prx2025, lapointe+prm2025}, and descriptor density of states (D-DOS) approaches use the internal representations of MLIPs as collective variables for model-agnostic free-energy estimation~\cite{swinburne+natcom2025}. 

Similar ideas are relevant at coarse resolution: machine-learned coarse-grained potentials define differentiable thermodynamic models that can be combined with free-energy perturbation to compute free energy differences such as mutation-based changes in folding free energies~\cite{charron2025}. These examples suggest that future benchmarks should evaluate not only force and energy errors, but also the stability of biased simulations, uncertainty under extrapolative sampling, and the accuracy of the final thermodynamic or kinetic observable.

\subsection{Generative samplers: Boltzmann generators and emulators}

Generative ML aims to bypass the timescale problem by drawing independent samples directly from a target distribution~\cite{Olsson2026}. One class of generative samplers are Boltzmann generators which combine generative models and statistical mechanical reweighting or Monte-Carlo acceptance to sample from a given equilibrium distribution of the form $\exp{-u(x)}$~\cite{noe+19science}. 
This approach was first demonstrated to reproduce equilibrium ensembles of small condensed matter systems and a small protein using normalizing flows~\cite{noe+19science}; later work extended the approach to other classes of systems, including Lennard-Jones clusters~\cite{pmlr-v119-kohler20a} and crystals~\cite{10.1088/2632-2153/ac6b16,Schebek2026}, studied its temperature dependence~\cite{Dibak2022,Moqvist2025},
and introduced flow-matching or diffusion models and annealing-inspired training~\cite{VigueraDiez2024,Lewis2025BioEmu,joshi2025allatom}. This versatile class of models has also been used in simulations to accelerate the convergence of free energy estimators~\cite{wirnsberger2020targeted,schebek2026assessing} or as proposals in more traditional Markov chain Monte Carlo settings~\cite{gabrie2022adaptive}. 

A limitation of Boltzmann generators is that they effectively generate the entire molecular structure in a single, global Monte Carlo move, and the requirement to match the proposal and target density with high precision practically limits their application to a few thousand degrees of freedom.
Alternatively, Boltzmann emulators such as AlphaFlow~\cite{Jing2024AlphaFlow} or BioEmu~\cite{Lewis2025BioEmu} are diffusion or flow models trained directly to generate ensembles using approximate equilibrium data from MD simulations or other sources. These models generally cannot be reweighted to a prescribed microscopic target density, so the task of learning an MLIP that defines the microscopic energy and the task of efficiently sampling from that energy remain separate. Recent approaches have attempted to bridge this gap by learning energy functions and samplers that are mutually consistent, or by deriving diffusion-model scores from a learned energy under constraints that match the diffusion-model distribution to the energy-induced distribution~\cite{arts2023onediffusionmodelsforce,diez2025boltzmann,Plainer2025ConsistentSampling}.

\subsection{Force-free integrators}
\label{sec:force-free}

A more recent line of work folds force prediction and time integration into a single learned evolution map~\cite{schr+23nips,bigi+25nips,Diez2026,Antoniadis2026,thie+26nmi,ripken2026learning,bigi+26prl}. By predicting finite-time updates directly, these models aim to overcome the small-step restriction imposed by high-frequency modes, making longer simulation times and larger molecular systems more accessible than with conventional MD. Conceptually, they learn a surrogate for the flow map of the dynamics, amortizing the effect of many small integration steps into one learned update. This approach changes the nature of the approximation problem: by absorbing time integration into the learned evolution map, it shifts the control of integration errors from the numerical solver into the model itself, making long-time accuracy, stability, conservation laws, and transferability central open questions.

For time steps beyond tens of femtoseconds, time integration becomes stochastic as the conditional transition density $p_\tau(x_{t+\tau}\mid x_t)$ must be sampled in order to represent molecular thermodynamics and kinetics. This leads to generative approaches for time propagation~\cite{Klein2023Timewarp,diez2025boltzmann, Antoniadis2026,schr+23nips}. Such models go beyond pure thermodynamic quantities as enabled by Boltzmann generators and emulators, and instead aim at preserving kinetic properties, such as transition rates, time-correlation functions and kinetic experimental observables, while significantly enhancing sampling beyond direct MD simulation.

\section{State of the art}
\label{sec:sota}

\begin{figure}
    \centering
    \includegraphics[width=\linewidth]{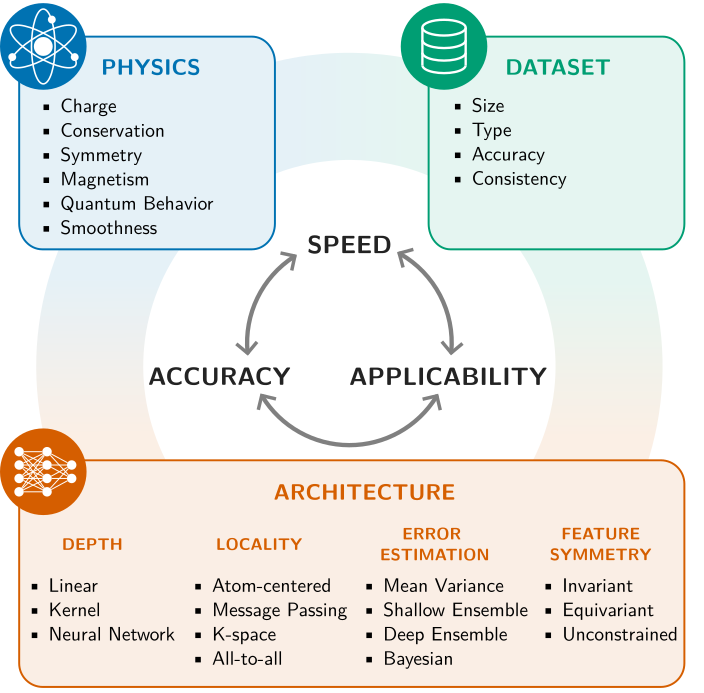}
    \caption{A schematic overview of the different elements determining the nature and the performance of an atomistic ML model.}
    \label{fig:sota}
\end{figure}

The evolution of ML models, in this field as in others, is driven by the need to maximize their performance---in terms of the sometimes conflicting goals of accuracy, speed, and breadth of applicability to classes of materials and simulation use cases.
There are many different possible ways to look at the ingredients that make up a model, but for the purpose of this roadmap we find it useful to discuss the model nature in terms of the physics that it is designed to incorporate, the mathematical architecture, and the type of data that are used (Fig.~\ref{fig:sota}). These three axes are obviously coupled (for instance, a symmetry-constrained architecture will ensure that a model makes predictions consistent with physical symmetries) but not in a trivial way (an unconstrained model can be symmetric to a very high degree if trained with data augmentation). 

\subsection{Physics}

Surrogate models of microscopic interactions must be able to describe the underlying physics, which serves at the same time as a consistency constraint and as inspiration for the mathematical structure of the model architecture.
First, physical considerations imply that the structure-property mapping must obey some general constraints, such as invariance or equivariance with respect to rigid translations and rotations, as well as some general properties such as smoothness with respect to atomic displacements, relationships between different quantities (e.g., forces being the negative gradient of the energy with respect to positions, which ensures that they behave as a conservative vector field) or (with the notable exception of electrostatics, polarization, and dispersion~\cite{khabibrakhmanov-2025}) nearsightedness of interactions~\cite{prod-kohn05pnas}.
More generally, the construction of an atomistic model benefits from being aware of the specific physical processes it should describe: Coulomb interactions, charge transfer, magnetism, quantum delocalization, electronic excitation, all imply a physical form of the interatomic interactions that can be translated into an appropriate parameterization of the model (such as including magnetic moments on atoms to describe magnetic interactions).
Note that it is not necessary for the model to take the physical form too literally. For example, electrostatics can be modeled by learning atomic charges explicitly~\cite{ghas+15prb,ko+21nc}, but also by building descriptors that can capture the appropriate asymptotic behavior of Coulomb interactions~\cite{gris-ceri19jcp,gris+21cs}, by learning atomic charges indirectly using only energy and forces as targets~\cite{Cheng_2025}, or simply by having charge-like atomic features that can be used to transfer information at long distance using a reciprocal-space electrostatics-like kernel~\cite{kosm+23icml,rumi+26tmlr}.

\subsection{Model architectures}
\label{sec:models}

The physical considerations above are ultimately reflected in the choice of model architecture. For example, to balance exploitation of locality with the ability to incorporate intermediate and long-range interactions, models have used architectures based on atom-centered descriptors~\cite{behl11jcp,bart+13prb,musi+21cr}, that describe the relative position of atoms within a set cutoff.
Message passing~\cite{gilm+17icml,schu+17ncomm,schu+18jcp} extends the effective receptive field of a per-atom prediction beyond the explicit cutoff~\cite{niga+22jcp2,bazt+22ncomm}.
As discussed above, genuine long-range information transfer can be achieved by manipulating structural descriptors in reciprocal space, explicitly or implicitly reproducing the functional form of Coulomb interactions.
It can also be achieved, in a less transparently physics-motivated manner, using all-to-all transformers~\cite{qu2026recipescalableattentionbasedmlips, kreiman2025transformersdiscovermolecularstructure}, linear scaling attention~\cite{frank2026machine} and virtual nodes~\cite{caruso2026}.

Architectures can also be distinguished by the complexity and depth of the functional form they are based on.
Shallow models retain a clear \emph{raison d'être}. Linear, non-linear and kernel approaches remain competitive and important for small data, fast inference, and distillation, especially when working on a well-defined chemical or configurational space~\cite{bart+10prl,P6721,shap16mms,chmi+17sa,chmi+18nc,deri+21cr}.
Modern MLIP architectures, especially when designed to cover large portions of chemical space, are dominated by graph neural networks acting on a neighbor list, with element embeddings that scale gracefully to chemically diverse training data.
While most current architectures employ a fixed computational depth, recent implicit architectures introduce adaptive depth, allowing the amount of computation to vary between atomic configurations~\cite{maeß2026implicitmachinelearningforce}.
Within the GNN family, a clear trend has been the integration of equivariance with respect to rotations and inversions.
Models that propagate equivariant features, typically \textit{via} tensor products on the irreducible representations of $\mathrm{SO}(3)$ and $\mathrm{SO}(2)$, have become very popular in the past few years~\cite{bazt+22ncomm,bata+22nips,musa+23ncomm,Batatia2025,frank2024euclidean,pmlr-v202-passaro23a,fu2025learningsmoothexpressiveinteratomic,li2026dpa4pushingaccuracycostfrontier}. High body-order representations~\cite{glie+18prb,will+19jcp,drau19prb,niga+22jcp2} and equivariant message passing belong to a family that includes the Atomic Cluster Expansion~\cite{drau19prb,duss+22jcp, darbyTRACE2023} and its neural relatives~\cite{bata+22nips,Batatia2025,P6787} that establishes formal completeness of local and semi-local representations~\cite{drau19prb,duss+22jcp,P6787}. At the same time, irreducible representations are not required for universality---invariant constructions built from inner products are already universal approximators of O(3)-vectorial~\cite{vill+21arxiv} and general equivariant~\cite{domi+25jcp} functions, and similar arguments apply to local message-passing architectures~\cite{klic+21arxiv}---so the value of equivariant, high-body-order architectures lies in inductive bias and data efficiency rather than in expressiveness in the limit.

At the other end of the spectrum, the need to incorporate symmetries into the structure of the model is increasingly being challenged: \emph{unconstrained} models that learn approximate equivariance from data, transformer-based potentials that relax the explicit graph locality assumption, and hybrid forms that mix the two have shown that strict symmetry adaptation, while elegant and data-efficient, is not always a hard requirement at the scale of modern training sets~\cite{unke2021spookynet,pozd-ceri23nips,mazi+25ncomm,qu2024the, qu2026recipescalableattentionbasedmlips, rhodes2025orbv3atomisticsimulationscale, kreiman2025transformersdiscovermolecularstructure}.
Recent work on plain transformers trained directly on Cartesian coordinates, with no graph and no built-in symmetry, has shown that locality and approximate equivariance can be recovered from data alone, and that empirical neural scaling laws of the kind familiar from language and vision apply also to atomistic models~\cite{kreiman2025transformersdiscovermolecularstructure,eissler2026simple}. 
The trade-off between bias and capacity is one of the more interesting open questions of current model design, and it intersects directly with the discussion of how to test models (Sec.~\ref{sec:assess}): emergent physical constraints, when they appear, may behave differently from hard-coded ones in out-of-distribution regimes, and require their own validation. 

Another architectural choice that clearly reflects the trade-off between exact compliance with physical constraints and a data-centric philosophy is that between conservative and direct force prediction. Conservative forces, obtained as gradients of an energy model, guarantee energy conservation in symplectic integrators, an essential property for thermodynamic sampling and traditionally considered as a quality measure of MD simulations~\cite{chmi+17sa}. Direct-force models avoid a gradient computation by predicting forces independently from energies~\cite{klic+21arxiv,neumann2024,bigi+25icml}, resulting in two- to three-fold faster inference at the cost of energy conservation. Where energy conservation remains desirable in the final model, the speed of direct force prediction can still be exploited: to reduce training cost during a pre-training stage~\cite{fu2025learningsmoothexpressiveinteratomic,bigi+25icml}, or to accelerate molecular dynamics through multiple-time-stepping schemes~\cite{tuck+92jcp} that use direct forces for the inner steps and conservative forces for the outer ones~\cite{bigi+25icml}.
The overall suitability of this design choice depends on the application, and several recent models offer both modes~\cite{mazi+25ncomm,rhodes2025orbv3atomisticsimulationscale,malosso2026high}.

\subsection{Data}
\label{sec:data}

The growth of training datasets has been one of the most visible trends of the past five years. Materials-focused datasets such as the OQMD~\cite{Kirklin2015} and Materials-Project-derived corpora~\cite{MP2013,MP2025}, OMat24~\cite{barrosoluque2026openmaterials2024omat24}, MPtrj~\cite{deng2023chgnet}, Alexandria~\cite{schmidt2023machine}, OpenLAM~\cite{peng2025openlam}, MAD~\cite{mazi+25sd}, MatterSim~\cite{yang2024mattersimdeeplearningatomistic}, MDR~\cite{koker2026pftphononfinetuningmachine}, GNoME-derived sets~\cite{merc+23nature}, SMAX~\cite{SMAX26} and others have pushed the count of structures with first-principles labels into the hundreds of millions. Molecular datasets such as QM7~\cite{rupp+12prl,mont+12nips}, QM9~\cite{rama+14sd}, ANI-1x~\cite{smit+17cs}, QM7-X~\cite{Hoja2021}, MD17/22~\cite{chmi+17sa, chmiela2023accurate}, SPICE~\cite{east+23sd}, Aquamarine~\cite{aqm}, and the GEMS~\cite{Unke2024}, OMol25~\cite{levine2026openmolecules2025omol25}, OPoly26~\cite{levine2026openpolymers2026opoly26}, QCML~\cite{qcml}, and QCell~\cite{Kabylda2026} datasets play an analogous role for organic chemistry and biomolecular fragments. 
Even where dataset sizes look comparable across efforts, the underlying first-principles settings are not. A dataset built with a particular pseudopotential, exchange–correlation functional, dispersion correction, plane-wave cutoff and other supporting grid densities, k-point mesh, and smearing is internally consistent but not necessarily compatible with another dataset built with different but equally reasonable choices. 
Magnetism, spin polarization, and Hubbard $+U$ values are particularly problematic~\cite{Warford_2026}: the same system can yield qualitatively different reference forces depending on whether collinear, non-collinear, or no spin treatment is used. 
The accuracy of DFT itself varies dramatically across the periodic table and across bonding regimes, in a way that is well known to electronic-structure practitioners but poorly accounted for in many ML pipelines.

These compatibility problems bias models in ways that are hard to detect downstream: a dataset that is internally consistent will validate well against itself even if it is systematically wrong, and the resulting model will inherit those errors silently. Foundation potentials inherit the same constraint: a model trained on a corpus built with a given functional and pseudopotential set is, in effect, a foundation model \emph{for that level of theory}, and porting it to a different reference (a different functional, an all-electron treatment, or a higher-accuracy method) requires either explicit fine-tuning on data at the new level~\cite{bata+25jcp, Kaur2025} or multi-fidelity strategies~\cite{rama+15jctc,sun-saut19jctc,smith2020ani1ccx,Dral2020hml, zhang2025graphneuralnetworkera, shoghi2024moleculesmaterialspretraininglarge, batatia2025crosslearningelectronicstructure,Gardner2025Understanding}.
Such multi-fidelity strategies explicitly model the offsets between methods
and can lead, for example, to a joint global potential energy surface based on data from DFT and correlated
wavefunction methods~\cite{Dral2020hml}.
In principle such approaches enable the training of ML models from a ``dataset of opportunity''~\cite{Fisher2024multitask}
amalgamated from existing databases of heterogeneous electronic-structure calculations.
However, while these methods can benefit from data of different sources~\cite{liang2025nep89universalneuroevolutionpotential, shiota2024tamingmultidomainfidelitydata, wood2026umafamilyuniversalmodels,Kim2026}
(potentially even without imposing an accuracy order \textit{across} sources~\cite{Fisher2024multitask})
they still require the data sources themselves to be internally consistent.
As such, it remains key that individual datasets are properly labeled with
standardized metadata (functional, pseudopotential version, force-drift checks, sum-of-forces tests, k-point convergence)
and verified through automated workflow validation.

Moreover, as all multi-fidelity approaches require a ground-truth anchor point,
to which all lower fidelities are related by the employed statistical model,
community-curated reference subsets at higher levels of theory
are essential.

Recent additions to the high-accuracy electronic-structure toolbox are also driven by AI innovations. Neural density-functional theory keeps the Kohn--Sham variational structure, but replaces part of the handcrafted exchange--correlation approximation by a learned functional. For instance, it was shown that neural exchange--correlation functionals trained with fractional-charge and fractional-spin constraints can reduce major delocalization and static-correlation errors~\cite{Kirkpatrick2021DM21}. More recently, Skala, a neural exchange--correlation functional trained on high-accuracy wavefunction data, achieved broad improvements on molecular thermochemistry benchmarks while retaining the $O(N^3)$ scaling and semi-local cost profile of Kohn--Sham DFT~\cite{Luise2025Skala}. If these methods become robust across chemical regimes, they will become valuable data generators for the next generation of MLIPs: much more accurate than standard density functionals, but cheap enough to generate larger and more diverse reference datasets than with CCSD(T).

A very high-end AI-enabled electronic structure method is variational Monte Carlo (VMC) with neural-network wavefunctions~\cite{Carleo2017Solving,Hermann2020PauliNet,Pfau2020FermiNet}. In this approach, energies and observables are evaluated as stochastic expectation values over configurations sampled from the neural wavefunction, while its parameters are optimized variationally~\cite{Carleo2017Solving}. Two complementary formulations have been developed. In first quantization, antisymmetric neural wavefunctions defined directly in continuous electronic coordinates, including PauliNet, FermiNet, PsiFormer and neural backflow/message-passing constructions, have reached very high accuracy for atoms, molecules, and extended interacting-electron systems~\cite{Hermann2020PauliNet,Pfau2020FermiNet,VonGlehn2023PsiFormer,Pescia2024MessagePassing}. In second quantization, neural wavefunctions instead parameterize the amplitudes of electronic configurations in an orbital basis, providing a complementary route to \emph{ab initio} molecular Hamiltonians and strongly correlated electronic structure~\cite{Carleo2017Solving,Choo2020Fermionic,Barrett2022Autoregressive}. 

The two representations have different computational bottlenecks and offer complementary advantages depending on the system and basis. Recent architectures and implementations, including PsiFormer and Forward-Laplacian-based approaches, have improved accuracy and scaling~\cite{VonGlehn2023PsiFormer,Li2024ForwardLaplacian}. Geometry-dependent and transferable neural wavefunctions create opportunities to amortize calculations over potential-energy surfaces rather than isolated geometries, and multi-state formulations extend this direction to excited states~\cite{scherbela2022solving,Gao2022PES,Gao2023Generalizing,Entwistle2023ExcitedVMC,Pfau2024ExcitedStates,rende2025foundation,Foster2025Orbformer}. Neural-wavefunction approaches are also being extended beyond the Born--Oppenheimer approximation to coupled electron--nuclear quantum systems, as well as to real-time many-electron dynamics~\cite{Linteau2026Hydrogen,Nys2024AbInitioDynamics}. Despite a polynomial system-size scaling~\cite{Pfau2020FermiNet,Schaetzle2023DeepQMC}, a very large prefactor makes these calculations very expensive, limiting routine applications to a few heavy atoms, and requiring effective core potentials for metals~\cite{Li_2022,Qian2022VMCForce}. Nevertheless, neural VMC is becoming a plausible source of high-end reference and fine-tuning data for strongly correlated chemistry, complementing DFT, coupled-cluster, multireference quantum chemistry, and emerging neural-DFT functionals.

A persistent split runs through the data landscape: materials-oriented datasets and molecule-oriented datasets have developed largely independently, and the energy and structural resolution that practitioners consider acceptable on each side differs significantly. The electronic-structure approximations themselves diverge: plane-wave pseudopotential DFT for periodic systems, hybrid functionals and post-Hartree–Fock methods for molecules. Training a single model across both worlds is increasingly common, but the incompatibilities in the underlying data are usually papered over rather than resolved.
Recent dataset building efforts emphasize that simply scaling up existing datasets following the same logic as high-throughput searches is not necessarily the most efficient approach, and have demonstrated that smaller datasets (with fewer than 1M structures) can be competitive as long as they are carefully selected to be representative of all classes of the materials and molecules domain, across many energy scales, and target highly converged, internally consistent, and/or higher-level-of-theory electronic-structure settings~\cite{matpes2025,mazi+25sd}.
This is one of the areas where coordinated community effort would have the highest leverage (Sec.~\ref{sec:strategy}): coordination would allow groups to generate datasets that are \emph{at the same time} diverse, well-curated, and high-accuracy, and larger than can be produced by individual groups. 

\subsection{What is now possible}
\label{sec:applications}

The applications enabled by current MLIPs are too numerous to do justice to in this section, and we list only a few representative directions~\cite{keith2021combining}. Since MLIPs for complex systems were first developed in the field of materials~\cite{behl-parr07prl,P1875}, large-scale and long-time simulations of amorphous and disordered systems, including glasses, liquid–solid interfaces, and complex defects such as grain boundaries and dislocations, are now routine~\cite{P3222,bart+18prx,caro+18cm,P4988,bern+19acie, allera+natcom2025, zhong+prx2025}.
Solid-state electrolytes, an area where MLIP-driven MD has reduced the cost of computing ionic conductivities by orders of magnitude~\cite{Deng2019Electrostatic,gigl+24cm,bohm2026}, and heterogeneous catalysis under realistic temperature, coverage, and pressure conditions~\cite{P7020}, are similar success stories. High-throughput discovery campaigns now use general-purpose potentials to scan thousands to millions of candidates for stability and for properties such as ionic conductivity~\cite{merc+23nature}. On the molecular side, water, as a key system for many questions in chemistry, has been a center of attention for a long time~\cite{P4556,P6721,symons2022application}. The controversy over liquid-water structure across exchange–correlation functionals~\cite{gill+16jcp} is now sufficiently well resolved that simulations can target experimentally accessible quantities with quantitative confidence~\cite{mont+24jcp}, in some cases at essentially coupled-cluster accuracy~\cite{P6360}. Matter under extreme pressure and temperature, long the preserve of shock experiments and diamond-anvil cells, has likewise come within reach of large-scale simulation: MLIPs now make it possible to follow pressure-induced phase transitions, melting curves, and the plastic and shear response of materials and minerals at conditions relevant to planetary interiors~\cite{chen+20nature,10.1145/3458817.3487400,SMAX26}. Single-molecule chemistry, including reactive pathways and excited-state dynamics, has been pushed forward by both kernel models~\cite{chmi+17sa,chmi+18nc} and neural surrogates~\cite{west-marq20mlst,west-maur21cs,unke2021spookynet}. Molecular materials and biomolecular systems remain harder, requiring efficient architectures and implementations~\cite{Kabylda2025}: still, classical force fields are competitive in terms of cost, and the timescales of interest are often beyond the reach of MLIP-driven MD without using the acceleration strategies discussed in Sec.~\ref{sec:beyond-props} or hybrid machine learning/molecular mechanics schemes~\cite{rufa2020towards,morado2025enhancing}. 

Several frontiers remain genuinely difficult. As discussed above, long-range electrostatics and dispersion were introduced in MLIPs long ago~\cite{P2391,P2962,10.1021/jp401225b}, but are not yet handled in a unified way; explicit charge-equilibration schemes~\cite{ghas+15prb,ko+21nc,rinaldi25}, latent long-range messages, and Ewald-like augmentations all coexist, and the suitability of a method is often determined by the physics of the system~\cite{ko+21nc,gao-rems22nc}. Self-consistent treatments are beginning to emerge~\cite{BaldwinSCF,BatatiaPOLAR1}, but the design space is large and remains to be explored.
There is not even consensus on \emph{when} an explicit long-range treatment matters, with some works finding very little practical impact on several bulk properties~\cite{Zaverkin_2026}.
Excited states, multiple spin states~\cite{P6057}, non-collinear magnetism~\cite{drau20prb,rinaldi24} and systems with strong static correlation push the limits of both the reference data and the model architectures. 
Furthermore, the indirect benefits of MLIPs are not fully exploited yet: for instance, even though MLIPs make it possible (and in a certain sense necessary) to simulate explicitly the quantum mechanical nature of light nuclei such as hydrogen, the sampling algorithms needed to model quantum nuclear effects~\cite{mark-ceri18nrc}, although they have been combined with ML~\cite{Musil2022,Zaporozhets2024}, are not yet broadly available inside end-to-end modeling tools. 

\section{Assessing the quality of models and data}
\label{sec:assess}

Constructing an MLIP should never be the final target of a study. It is a step in a longer toolchain that ends in the calculation of a thermodynamic property, a dynamical observable, or a structural prediction, sometimes based on combined information from millions of evaluations. A model that scores well on per-configuration energy or force RMSE of a fixed test set may still produce unphysical trajectories, fail to reproduce melting points or radial distribution functions, or break down in long simulations. Conversely, a model with an apparently mediocre RMSE may be perfectly adequate for the actual question being asked, because the errors average out or cancel between configurations. A further difficulty is that many properties, such as free energies, reaction rates, or transport coefficients, have no reference value to test against: unlike energies, forces, and stresses, they cannot be computed at the reference level of electronic structure theory at acceptable cost, and they emerge from the potential only together with a model of the dynamics and a finite compute budget, so a disagreement with experiments may stem from the potential, the dynamics, or simply too short a simulation. Assessing the quality of MLIPs, and of the data they are trained on, is therefore more complicated than simply validating a regression model, and the community is still developing the tools to perform this task.

\subsection{Beyond test-set error}

\begin{figure}
    \centering
    \includegraphics[width=\linewidth]{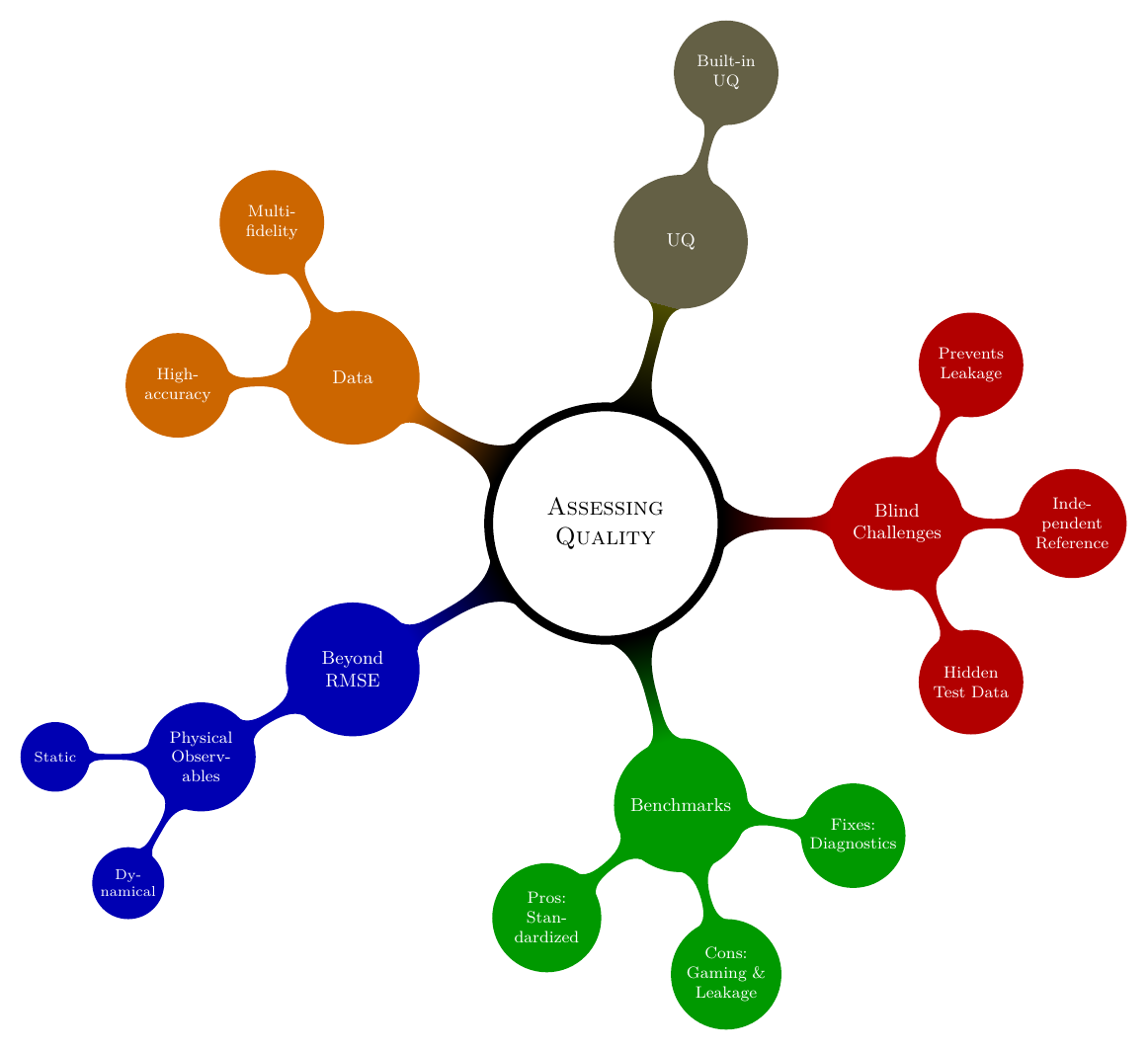}
    \caption{Thorough evaluation of MLIPs requires assessment beyond static error metrics on energies and forces.}
    \label{fig:assessment}
\end{figure}

The first issue is that the test-set RMSE is necessary but rarely sufficient as a quality measure. The distribution of configurations sampled by a thermalized molecular dynamics simulation is not the distribution of the training set; transferability away from equilibrium, smoothness of the potential energy surface, and stability under long integration are all properties that the training loss or the test set error do not directly measure~\cite{Poltavsky2025}.
At the very least, the distribution of errors should always be reported, as well as learning curves as a function of training-set size that provide information on model capacity~\cite{vonlilienfeld2020exploring,rupp+12prl}.
The community has accumulated a body of practical experience showing that ``beyond RMSE'' tests or downstream tasks, including stability checks under high-temperature or near-singular geometries, computation of static observables (radial distribution functions, mean square displacements, elastic constants, defect formation energies), dynamical observables (diffusion coefficients, melting temperatures, vibrational spectra), and macroscopic thermodynamic targets, are essential complements to the standard error metrics~\cite{10.1063/5.0139611,isamura2026unprecedented} (cf. Fig.~\ref{fig:assessment}).
These are also the tests that bring out the relevance of \emph{distributions} of errors rather than their averages: small parts of a system, such as a reactive site, a defect, a transition state, or an interface, can dominate the physics while contributing negligibly to a global RMSE.

A complementary observation is that benchmarks on raw labels are only as good as the labels themselves. Reference energies, forces, and stresses inherit the systematic errors of the underlying first-principles method, so an ``error of 1\,meV/atom'' against DFT does not translate directly into an error of 1\,meV/atom against experiment or against a higher-level quantum-chemistry reference. Benchmarks should therefore report the spread across reasonable DFT settings alongside the headline error, annotate data with DFT uncertainty estimates (see Sec.~\ref{sec:uncertainty}), and, where a higher level of theory is available, quote the model error against that reference as well; only then does the reported fidelity reflect the model rather than an arbitrary choice of reference.

\subsection{Benchmarks: useful, but not as goals in themselves}

Benchmarks have been a force for good in atomistic ML. They have improved accessibility, lowered entry barriers, and produced visible prediction accuracy gains across the field~\cite{Butler2024Setting}. Established platforms such as Matbench Discovery~\cite{riebesell2025}, MLIP Arena~\cite{chiangMLIPArenaAdvancing2025}, ML-PEG~\cite{kasoar_2025}, LAMBench~\cite{Peng2026}, the Fairchem leaderboard~\cite{levine2026openmolecules2025omol25} and others now offer standardized comparisons across a substantial subset of the model landscape with various degrees of diversity across the validation sets. The accumulated evidence, however, also suggests that benchmarks become problematic when they turn from a measurement tool into the actual goal of method development, e.g., a pure leaderboard rather than a tool of understanding. Once a benchmark becomes a target for funding, publication, or career visibility, its incentive structure pushes models towards designs that score well on it, sometimes at the cost of performance on out-of-distribution problems that matter more in practice (see, e.g., the Clever Hans effect~\cite{lapuschkin2019unmasking,kauffmann2025explainable}, in which models predict correctly but for the wrong reason). A subtler problem is that benchmarks erode as the field's data grows: the more widely a benchmark is adopted, the more likely its test structures are to migrate, usually inadvertently, into the training sets of later models, a form of data leakage that silently inflates apparent accuracy.
This failure mode is well documented in other ML communities, most prominently in the contamination of large-language-model evaluations~\cite{Sainz2023NLP}, and atomistic ML is not immune.
It is a further reason for benchmarks to carry an explicit life-cycle, with periodic refreshing or retirement.

A handful of design principles can keep benchmarks useful and honest: 
\begin{itemize}
    \item they should include physically meaningful diagnostics alongside raw-label errors
    \item they should be clear about which settings of the reference method were used and why
    \item they should make a best-effort estimate of the intrinsic noise and uncertainty in the benchmark labels
    \item they should explicitly discuss the limitations of the reference method used to generate the targets
    \item they should be neutral with respect to method development, in the sense that the benchmark maintainers and the model developers should not be the same group
    \item they should also provide a standardized, unbiased assessment of the computational cost
    \item they should have an explicit life-cycle, including a way to retire benchmarks that have outlived their usefulness
\end{itemize}
When a benchmark targets an emergent observable (a phase boundary, a reaction free energy, \ldots) rather than a raw microscopic label, there should be strict accounting of the simulation protocol and costs, preferably over a range of different systems, to gauge both transferability and efficiency. Benchmarks should also be openly available to all stakeholders---including experimentalists, engineers, and domain scientists beyond the ML community---so that their scope and relevance can be collectively shaped. A portfolio of complementary tests is preferable to a single number that ranks all models: it is less convenient, but it rewards diversity of approach and makes it harder to game any one test.

\subsection{Community one-time challenges as a complement to benchmarks}\label{sec:onetime}

A different incentive structure emerges from community one-time challenges. Inspired by CASP in protein structure prediction~\cite{yuan2026casp16} and the CSP blind test~\cite{hunnisett2024seventh} in crystal structure prediction, a challenge periodically asks the community to predict outcomes for problems that are blind to the participants. Challenges are hard to organize and resource-intensive, particularly when they require the acquisition of new experimental measurements as reference, but they have several attractive properties that benchmarks lack: by design, the test set cannot be in the training set (although data leakage can still take place when the new data points are similar); the reference is generated independently, often after the predictions are submitted; and the visibility benefits accrue to the participants who genuinely solve a problem of interest, rather than those who optimize against a fixed leaderboard. 
They are, moreover, the natural way to assess properties for which no affordable computational reference exists, since an independent experimental measurement can supply the ultimate grounding in physical reality that a calculation cannot. We see one-time challenges as a complement to, not a replacement for, benchmarks: each addresses a different question. Indeed, blind challenges often serve to point to areas where current benchmarks fail to be predictive. A particularly attractive variant has the winner of one challenge organize the next, with the constraint that they cannot themselves participate, naturally rotating responsibility while keeping continuity. A challenge with the described characteristics would enable fair evaluation of methodologies while at the same time encouraging a healthy competitive environment among the current participants and inspiring contributions from new actors. Initiatives in this direction are only starting to emerge in the community, e.g., the LAM Crystal Philately competition~\cite{peng2025openlam}.

\subsection{Uncertainty and validity}
\label{sec:uncertainty}
Whether through benchmarks or challenges, end users ultimately need to know not only ``which model is most accurate or efficient or both on average'', but ``what is the expected accuracy of a given model for my system, and how do I know it has not silently broken down''. Built-in uncertainty quantification is a key part of the answer. Ensemble approaches~\cite{P5814}, deep evidential regression~\cite{1910.02600}, last-layer prediction rigidity~\cite{bigi+24mlst}, methods to partition uncertainty into basis/representation, local/data-density, and model-form contributions and various calibration schemes are all available with varying overhead~\cite{tran+20mlst,kell-ceri24mlst}; what is missing is an established practice in which uncertainty estimates accompany every published model, are themselves validated, and are reported through interfaces standard enough for downstream tools to consume. Calibration of the uncertainty is itself an open methodological question, but the community can and should converge on a baseline practice now and refine it later. Consideration of uncertainties due to model mis-specification~\cite{Swinburne2024} as well as approaches based on ensuring internal consistency of models~\cite{Qian2026} should complement reliance on calibration to hold-out validation sets.
The accuracy of the underlying first-principles method must enter the same conversation: a 5\,meV/atom uncertainty in the model might be viewed as meaningless if the reference DFT itself is uncertain at the level of 50\,meV/atom for the property of interest, and even for the same functional, results depend on numerical choices such as the pseudopotential~\cite{leja+16science}, although errors of MLIPs and DFT are often of a different nature, with the latter being more systematic and therefore leading to better cancellation of errors for derived properties. At the same time, the high flexibility of MLIPs can lead to essentially unusable models for practical simulations beyond some level of pointwise error. 
Ultimately, one should aim for a holistic error metric that provides the error against experiments. This is not without challenges, since experimental data are abundant for macroscopic observables but scarce for microscopic ones, and are themselves subject to errors; moreover, the error with respect to experiment depends on the accuracy of the MLIP, the accuracy of the reference method, and, for properties extracted from simulation, the dynamics and the amount of sampling; these contributions must be disentangled to know what to improve.

In this regard, the ongoing developments towards the practical estimation of DFT model and
numerical errors are worth highlighting. This includes the development of probabilistic
DFT models, such as the BEEF family of models~\cite{Mortensen2005beef,Christensen2020uqmmm}
and uncertainty-aware functional distributions (UAFD)~\cite{hans+25prb}, which replace a single choice of DFT parameters by calibrated
distributions, encoding experimental uncertainty. In combination with
differentiable DFT codes~\cite{Kasim2022,pyscfad,MCasares2024,DFTKpaper}
these parameter distributions can now be systematically and efficiently
propagated to arbitrary quantities of interest
such as DFT-optimized structures or forces~\cite{Schmitz2025addfpt}.
Using MLIPs to inexpensively reproduce calibrated DFT ensembles might provide affordable end-to-end uncertainty quantification that reports directly on the experimental error, rather than on the error relative to a first-principles baseline of unknown accuracy~\cite{kell+arxiv}.
Furthermore, recent mathematical advances in the estimation of the plane-wave basis set error~\cite{Cances2022error}
enable the efficient estimation of basis set errors in DFT forces and other quantities
of interest~\cite{Schmitz2025addfpt}.
As these approaches continue to mature, considering such error estimates
during MLIP training becomes viable and offers hope to systematically quantify the effect
of changing DFT parameters (basis set cutoff energy, pseudopotential, Hubbard-$U$ parameter, etc.)
on MLIP predictions in the future.

Finally, since MLIPs are trained on electronic-structure calculations rather than on experimental measurements~\cite{kulichenko2021rise}, their accuracy against experiment is ultimately bounded by that of the reference method: even a model with zero generalization error reproduces the self-interaction and delocalization errors of the functional it was trained on. Crossing this ``quantum ceiling'' requires training data from higher levels of theory, such as coupled-cluster calculations, whose cost precludes the routine generation of large datasets. Multi-fidelity and $\Delta$-learning strategies~\cite{vonlilienfeld2020exploring} (Sec.~\ref{sec:data}) and hybrid training against experimental targets (Sec.~\ref{sec:cg}) are the most promising routes to bridge this gap.

\section{A challenge of software and hardware}
\label{sec:swhw}

The advances of past years have been made possible by an underlying revolution in software and hardware that the atomistic community largely did not drive. The dominant tooling, PyTorch~\cite{Paszke2019} and JAX~\cite{jax2018github} above all, was designed for deep-learning workloads in industry; the dominant accelerators are GPUs and increasingly specialized artificial intelligence (AI) hardware whose roadmaps are set by a small number of vendors. This has been an enormous boost, and it has strongly influenced the evolution of the computational ecosystem (software, hardware, infrastructure). It also creates fragility: the lower layers of the stack evolve at a pace that the typical scientific code base cannot follow without continuous investment, and key low-level components (from vendors) are not always open source.

Historically, atomistic simulations have relied on classical force fields implemented in C++ or Fortran MD engines designed around spatial domain decomposition, whereas modern ML architectures are developed predominantly within Python-centric frameworks such as PyTorch and JAX. Bridging the two requires the model's structural representations and neighbor-list routines to execute natively on accelerators and within domain-decomposed MD codes~\cite{park2024scalableparallelalgorithmgraph}, which is particularly demanding for equivariant message-passing networks, where tensor products on irreducible representations must be evaluated on the fly~\cite{tan2025highperformancetraininginferencedeep}. Modern implementations can also execute automated active-learning loops directly on distributed HPC systems, flagging extrapolative configurations during long production runs~\cite{novikov2021mlip,lysogorskiy2023active}.
Assembling these closed loops, however, still demands expert knowledge across all involved domains, and dedicated workflow frameworks aim to reduce the barrier to entry by packaging recurring operations as reusable, modular building blocks that can be shared across groups and projects~\cite{zillsCollaborationMachineLearnedPotentials2024,gelzinyteWflPythonToolkit2023,huberAiiDA10Scalable2020,ganoseAtomate2ModularWorkflows2025,menon24}. Looking further ahead, agentic interfaces may orchestrate computational workflows directly~\cite{phamChemGraphAgenticFramework2026} and coordinate entire parts of the research process~\cite{zouAgenteAutonomousAgent2025}, which could make high-throughput active-learning pipelines accessible to non-specialists.

\subsection{Anatomy of the stack}

\begin{figure}
\centering
\includegraphics[width=\columnwidth]{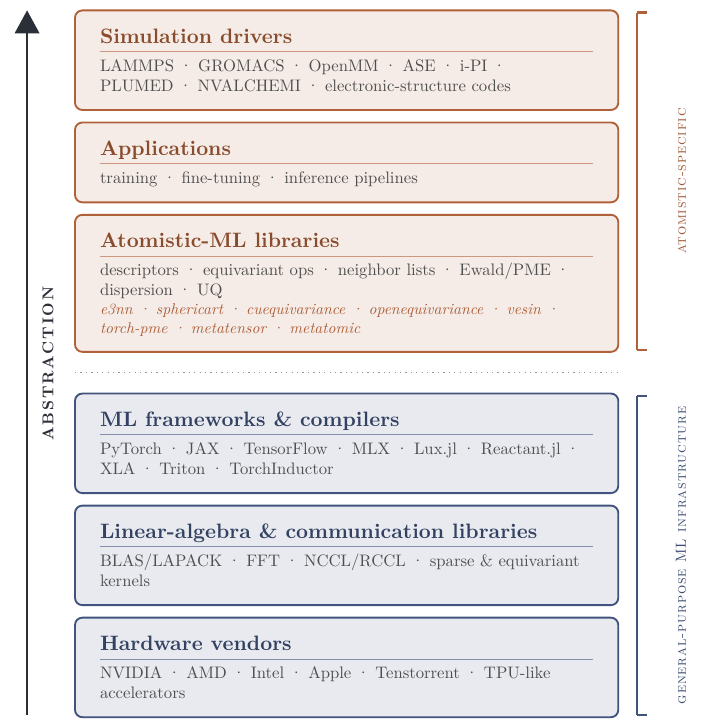}
\caption{The atomistic-ML software and hardware stack, organized as layers of increasing abstraction.}
\label{fig:stack}
\end{figure}

A useful way to organize the discussion is to think of the atomistic-ML stack as a small number of layers with very different characteristics (Fig.~\ref{fig:stack}). At the bottom sit the hardware vendors and the linear algebra, tensor-operation, and communication libraries (BLAS/LAPACK~\cite{lawson1979,rice1988,dongarra1990,dongarra2002,angerson1990}, FFT~\cite{frigo2004}, NCCL~\cite{nvidia_nccl}/RCCL~\cite{rocm_systems}, kernel libraries for sparse and equivariant operations). Above them, ML frameworks (such as PyTorch, JAX, TensorFlow~\cite{tensorflow2015-whitepaper}, MLX~\cite{mlx2023}, Lux.jl~\cite{pal2023lux}, Reactant.jl~\cite{reactant_jl} and a long tail of compilers and intermediate representations such as XLA~\cite{openxla_xla}, Triton~\cite{tillet2019}, TorchInductor~\cite{ansel2024pytorch2}) provide automatic differentiation, kernel scheduling, and a programming surface. On top of these, atomistic-ML libraries provide higher-level building blocks: descriptors, equivariant operations, neighbor lists, dispersion corrections, Ewald summation and particle–mesh routines, and uncertainty estimation. Applications then build on these to define training pipelines, fine-tuning workflows, and inference engines. Finally, simulation drivers, such as LAMMPS~\cite{plim95jcp}, GROMACS~\cite{abraham2015gromacs}, OpenMM~\cite{eastman2023openmm}, ASE~\cite{Larsen2017} (and a plethora of other workflows built on top of it), i-PI~\cite{ceri+14cpc}, PLUMED~\cite{bono+19nm}, NVALCHEMI~\cite{nvalchemi_toolkit,nvalchemi_toolkit_ops}, and the various electronic structure packages, embed the inference engines and orchestrate molecular dynamics, geometry optimization, enhanced sampling, data labeling and analysis.

The trouble is that this stack is currently fragmented in ways that hurt productivity and reproducibility. Each architecture tends to ship with its own training code, its own dataset format, its own neighbor-list implementation, and its own bindings to a subset of simulation drivers. Re-implementations of a LAMMPS or OpenMM interface, with subtle differences, are a source of repeated efforts and are often poorly structured to leverage typical ML performance strategies such as batching that are strong drivers of hardware and algorithmic development. Low-level operations that should be shared (equivariant tensor products, neighbor-list builders, dispersion corrections, Ewald summation) often are not, and the few cases where reusable libraries have emerged (e3nn~\cite{e3nn}, sphericart~\cite{bigi+23jcp}, cuEquivariance, OpenEquivariance~\cite{openeq}, FlashTP~\cite{lee25l}, vesin~\cite{bigi+26jcp}, torch-pme~\cite{loch+25jcp}, matscipy~\cite{Grigorev2024-dp}, metatensor~\cite{bigi+26jcp}) have not always resulted in standardized adoption.

The complexity and heterogeneity of the stack increase even further if one considers efforts at the intersection between ML and electronic-structure theory: the large memory footprint of calculations, the diversity of underlying formalisms (plane waves, Gaussian orbitals, numerical atomic orbitals, etc.) and the need for high levels of parallelism make it even harder to conceive a modular design and the definition of interfaces with ML libraries.
That said, there are recent examples of autodifferentiable quantum chemistry codes~\cite{Kasim2022,pyscfad,MCasares2024,DFTKpaper,Schmitz2025addfpt} relying on similar software frameworks as the traditional atomistic-ML stack,
which are promising for such efforts.

\subsection{Reasons for monolithic codes, and reasons against}

There are reasons why monolithic codes have thrived. Tight integration of the components allows aggressive optimization, simpler governance with clear responsibilities, and a single design philosophy. For graduate students and small groups, a self-contained package gives a clear publishable artifact and avoids the overhead of negotiating interfaces with collaborators. Modular ecosystems have their own pathologies: dependency hell, version pinning that breaks downstream tools, the need for coordination across different components, the diffusion of responsibility when a bug spans two libraries, and the difficulty of tracking citations and giving credit for low-level building blocks. We do not argue for a single architecture, nor for a forced consolidation. We argue that the field has reached a degree of maturity where the costs of fragmentation now outweigh the benefits of unhindered independent development, and where modest investment in shared interfaces would unlock disproportionate gains.
This balance is shifting further as AI coding agents mature.
As the generation and maintenance of bindings, glue code, and interface boilerplate---including chasing breaking changes in fast-moving upstream libraries---becomes increasingly automated, the implementation effort that once justified folding everything into a single tightly integrated codebase is becoming a commodity, lowering the barrier to assembling systems from independent, interoperable components.

A useful guiding principle, articulated during the workshop, is to separate \emph{algorithms}, \emph{reference implementations}, and \emph{performant implementations}. The community should agree on a small set of algorithmic primitives that matter (e.g., segmented sparse operations, equivariant tensor products, Ewald-like long-range routines, neighbor-list construction, shape-flexible loss functions). These should have clean reference implementations that are primarily easy to read and modify, and one or more performant implementations, possibly closed-source and vendor-supplied, that are interchangeable behind the same interface.
This pattern has worked well in numerical linear algebra and signal processing for decades: there is no fundamental reason it should not work here.

We remark that modern software frameworks and programming languages such as JAX, Julia or PyTorch can in fact
blur the distinction between a reference and a performant implementation,
in the sense that implementations employing those languages can remain hackable,
while still featuring production-grade efficiency~\cite{DFTKpaper,scho+21jsm,Greener2024molly,loch+25jcp}.
As a result, core computational routines typically remain accessible
and not hidden away in low-level kernels written in a separate language.

This is crucial for mathematical research, which requires both
the flexibility to explore ideas on well-controlled toy problems
\emph{and} the gradual upscaling of algorithms
towards the full atomistic modeling setting.
In this niche the Julia programming language
has grown considerably in popularity and has played a role
in the development of new sampling algorithms~\cite{Blassel2024},
of the MD engine Molly~\cite{Greener2024molly},
of algorithms for ACE model training in ACEpotentials.jl~\cite{Witt2023ace},
and of the aforementioned DFT error estimation strategies~\cite{Schmitz2025addfpt}
in the Density-Functional ToolKit~(DFTK)~\cite{DFTKpaper}.
For such efforts the key advantage of Julia
is that custom algorithms (including novel bottom-level linear algebra routines)
can be written in a high-level language with only minimal performance impact
while still fully integrating with other flagship features
such as GPU acceleration or differentiability.
This makes individual Julia tools competitive for many research tasks, but their integration with standard simulation ecosystems---a prerequisite for wider adoption---remains rudimentary.

\subsection{Interoperability and isomorphic interfaces}

Where standardization has emerged in our field, it has done so organically. The ASE Calculator interface, with its simple positions–species–cell to energy–forces–stress contract, has become the \emph{de facto} common surface for MLIPs over the past few years, despite never having been formally proposed as a standard, and despite the limitations associated with the Python framework. 
Useful precedents include LAMMPS's ML-IAP plugin, OpenKIM~\cite{tadmor2011potential} for classical force fields, the interfaces of the JuliaMolSim community~\footnote{For the AtomsCalculator interface of the JuliaMolSim community see \protect\url{https://juliamolsim.github.io/AtomsCalculators.jl/stable/interface}.} employed by most Julia-based atomistic simulation software,
as well as the metatomic interface that abstracts the MD engine away from the model~\cite{bigi+26jcp}. 
Looking forward, an explicit specification of an isomorphic calculator interface, language-agnostic and implementable in Python, Julia, C, C++, Fortran, Rust, or any future language, would be a modest investment with a large potential return for the community. The same logic applies to dataset formats: the field is converging on a small set of options (extended XYZ, HDF5, LMDB, Zarr, Parquet) and would benefit far more from agreement on a common ontology of fields than on a single binary representation. A naming convention that distinguishes ``Hirshfeld charges'' from ``Mulliken charges'', and that allows for new fields to be added without breaking old code, is the kind of work that pays for itself in the medium term.

Plugin-style architectures are a natural complement: a lean core, with heavyweight or domain-specific functionality in optional modules that can evolve and be deprecated independently. ASE itself is moving in this direction. A \emph{language-agnostic} calculator specification would naturally come with bindings in multiple languages and a stable C ABI that allows production codes (often written in Fortran or C++) to call into ML inference engines without depending on the entire Python ecosystem. Compiling models to language- and architecture-agnostic formats such as the Open Neural Network Exchange (ONNX) or StableHLO on MLIR (Multi-Level Intermediate Representation) has the potential to enable further interoperability. Yet modularity must be designed with a global view of the software stack: plugin boundaries should be chosen so as not to introduce performance bottlenecks, and architectural decisions at the interface level should be informed by end-to-end performance considerations rather than local convenience. By making interface code cheap to write, AI coding agents also raise the stakes for proper design and rigorous testing, which too often are second-class citizens in scientific software development.

\subsection{Hardware: leverage and lock-in}

The atomistic-ML community's relationship with hardware vendors is asymmetric. Most modern training and inference happens on NVIDIA GPUs, with AMD, Intel, Apple, Tenstorrent, and various TPU and TPU-like accelerators in supporting roles. The \emph{de facto} centrality of NVIDIA reflects the maturity of CUDA and the open-source ecosystem built around it; this also creates risks. The community has limited leverage to demand support for our specific operations from large vendors, but it has more leverage than it currently uses, and acting as a coordinated voice rather than as a collection of individual groups would help. 
Concretely, agreeing on a small set of operations that we collectively care about and engaging with vendors and HPC centers around those operations is far more effective than each architecture group negotiating its own kernel support. Vendor-optimized operations can also smooth the transition to new hardware, something that research groups with limited resources struggle to do. The field seems to be moving in this direction, and vendors are engaged in a constructive fashion. The fact that the community has also been actively developing fully open, hardware-agnostic domain libraries mitigates the risk of vendor lock-in.

Performance comparisons across architectures, hardware, and software stacks are notoriously hard. The same algorithm can vary by an order of magnitude in throughput depending on implementation details; reasonable benchmarks must specify hardware, drivers, compilers, batching strategy, and which parts of the pipeline are timed. Whole-MD throughput, including the simulation driver, is the most application-relevant metric and the one most prone to confounding factors. The community would benefit from a standardized harness, ideally maintained by an entity that does not also develop a competing model, that runs end-to-end MD throughput tests on a fixed set of hardware. A comprehensive definition of benchmarking protocols, ontologies and a publicly available database of results would be beneficial for the community and the whole HPC ecosystem, including technology providers and funding agencies. In this respect, the HPC community is the natural partner for the scientific community to move forward.

\section{Strategy for the community}
\label{sec:strategy}

The previous sections have surveyed where the field stands and what the key open problems are. We close with a small number of concrete actions that, in our view, would most improve the long-term health of the atomistic-ML ecosystem. None of them is technically novel; their value is in being adopted as community practice rather than left to individual goodwill.

\paragraph{Coordinate, do not consolidate.} The field benefits from a diversity of architectures, training pipelines, and software stacks. The right level of action is interoperability, not unification, also because the field is still rapidly evolving. It is simply too early to stop significant developments by narrowing down the ecosystem. In practice this means agreeing on a small set of language-agnostic interfaces (a calculator API, a dataset ontology, model-card metadata) and on a small set of low-level primitives that should be shared across architectures. Several of these already exist (ASE~\cite{Larsen2017}, metatensor and its derivatives~\cite{bigi+26jcp}, e3nn~\cite{e3nn}, sphericart~\cite{bigi+23jcp}, vesin~\cite{bigi+26jcp}, torch-pme~\cite{loch+25jcp}, NVALCHEMI, the JuliaMolSim interfaces); the action is to formalize their APIs in a collaborative fashion, document them, define their governance and resource their maintenance.

\paragraph{Datasets: depth, breadth, and consistency.} The community should prioritize the production of higher-fidelity reference data, especially for systems and properties where DFT is uncertain (strong correlation, transition metals, magnetism, non-covalent interactions, condensed-phase nuclear quantum effects). Sharing should be supported by metadata standards that record functional, pseudopotential version, k-point grid, spin treatment, and basic sanity-check quantities such as force drift; multi-fidelity strategies should be encouraged where they offer the most leverage. 
To make this enforceable rather than aspirational, the community should designate a small, public set of reference structures---a few hundred spanning the main bonding regimes---that every contributed dataset recomputes and reports at its own level of theory; the resulting fingerprints make electronic-structure settings comparable across datasets and expose silent incompatibilities before models are trained on them. 
Computing centers and data hosts can play a much larger role than they currently do, both as long-term repositories and as providers of CI infrastructure that runs sanity checks on contributed datasets, even though long-term support might still rely on individual engagement more than on institutional mechanisms.

\paragraph{Benchmarks and challenges.} Existing benchmarks should be embraced as part of the ecosystem, but supplemented with diverse, application-driven tests that go beyond label RMSE; with explicit reporting of the underlying reference settings; and with mechanisms to retire benchmarks that have outlived their purpose. 
Standardized assessment of training and inference cost would provide complementary information to validation accuracy, and drive development to minimize the energetic and environmental impact of the field.
In parallel, we propose a regular (e.g., biennial) series of blind challenges paired with a community conference, hosted by CECAM and run by a steering committee that rotates on the winner-organizes-the-next principle of Sec.~\ref{sec:onetime} and is resourced by sponsoring institutions. The same committee would act as custodian for the standards called for throughout this roadmap---the calculator API, the dataset ontology, model-card metadata, the shared low-level primitives, and the recommended UQ practices---ratifying a versioned release of each at every meeting and naming a maintainer of record, so that the recurring call to ``agree as a community'' resolves to a named venue and an explicit decision procedure.

\paragraph{Controlled model problems.} 
Important theoretical questions highlighted throughout this roadmap---such as when a model extrapolates, whether emergent constraints behave like hard-coded ones, and how representations and architectures limit attainable accuracy---can be sharpened, and sometimes settled, through carefully chosen model problems: reduced settings that isolate specific mechanisms, such as a known long-range tail, an isolated defect, or an analytically tractable energy landscape, in situations where the ground truth is known by construction. Combined with the theoretical tools developed around them, such problems can deepen our understanding of the fundamental capabilities and limitations of ML models, or provide a rigorous basis for substantiating or falsifying empirical claims.

\paragraph{Uncertainty quantification as default, not feature.} 
Every model intended for production use should ship with calibrated uncertainty estimates, validated on held-out data, and exposed through a standard interface of per-atom uncertainties and hooks for propagation to derived properties. Uncertainty must be reported alongside the accuracy of the reference method. A small set of community-recommended UQ approaches, with reference implementations, would lower the barrier to adoption.

\paragraph{Reward software and infrastructure work.} Sustainable shared infrastructure cannot be built on volunteer effort alone. Career paths for research software engineers, with progression comparable to academic ranks, are essential, and several countries have started to provide them; the community should reinforce that trend by giving software contributions visible academic credit. Citation conventions for low-level libraries (CITATION.cff, archival DOIs, explicit acknowledgment in publications) should be enforced as a matter of practice.

\paragraph{Engage traditional codes and vendors as partners.} The boundary between ``traditional'' simulation codes and modern ML-driven workflows is increasingly artificial. Electronic structure codes, sampling drivers, and ML inference engines need to interoperate at the level of in-memory data, not text files. Joint workshops, shared CI infrastructure, and joint hires with HPC centers and vendor research groups would accelerate this convergence. Vendor lock-in is a real risk; the antidote is not to refuse vendor support, but to ensure that vendor-supplied implementations sit behind community-defined interfaces that admit alternatives.

\paragraph{Stewardship.} Finally, this roadmap should be maintained as a living document rather than a snapshot: the steering committee proposed above should publish a versioned update of these recommendations after each meeting, tracking which actions have been delivered and which remain open. We invite the broader community to make use of the online and offline spaces that already exist.\footnote{An online community space is available at \protect\url{https://ml4atoms.org/slack}.}

The atomistic machine-learning ecosystem has outgrown its early, exploratory phase. Its scientific impact is now sufficient to justify the modest amount of coordination this roadmap recommends, and the cost of inaction, in duplicated effort, lock-in, and lost scientific opportunity, is rising. The opportunity is there. It will not stay open indefinitely.

\section*{Author contributions statement}

JB, MC, CC, GC, A-ME and AK organized the CECAM meeting and/or coordinated the on-site discussion, prepared a first draft consolidating the minutes of such discussion and finalized the manuscript.
MC coordinated the manuscript preparation and handled the editorial correspondence.
All other authors have read the manuscript, provided comments and suggested edits, and agree with the strategic vision laid out in the roadmap. AI was used to summarize workshop notes, and to prepare an early draft of the roadmap based on an extended outline.

\begin{acknowledgments}
We thank CECAM for hosting the workshop that originated this roadmap, as well as Psi-k and Achira for sponsoring it. 
We are grateful to all participants for the vigorous and generous discussion that shaped its contents, and the broader atomistic machine-learning community whose contributions, online and offline, kept informing the analysis after the workshop ended.

\end{acknowledgments}

\bibliographystyle{apsrmp4-1}

\end{document}